\documentclass[conference]{IEEEtran}

\usepackage{my_style}
\makeatletter
\def\ps@headings{
\let\@oddhead\@empty
\let\@evenhead\@empty
\def\@oddfoot{\@IEEEheaderstyle\hfil}%
\def\@evenfoot{\@IEEEheaderstyle\thepage\hfil\hbox{}}
}
\def\ps@IEEEtitlepagestyle{
\let\@oddhead\@empty
\let\@evenhead\@empty
\def\@oddfoot{\scriptsize Copyright © 2024 by Zeyad M.~Manaa; authors retain rights until potential transfer to IEEE upon publication in the proceedings of the 3rd IEEE Conference on Smart Mobility. \hfill {}}%
\let\@evenfoot\@empty
}
\makeatother

\begin{document}
\setlength{\textfloatsep}{5pt}

\title{Koopman-LQR Controller for Quadrotor UAVs from Data\\
\thanks{

This work was supported by the Interdisciplinary Research Center for Aviation and Space Exploration (IRC-ASE) at King Fahd University of Petroleum and Minerals under research project/grant INAE 2401.\\
\indent Z.~M.~Manaa, A.~M.~Abdallah, and S.~S.~A.~Ali are with the IRC-ASE, and the Department of Aerospace Engineering at King Fahd University for Petroleum and Minerals, Dhahran, 31261, Saudi Arabia.\\
\indent M.~A.~Abido is with the Interdisciplinary Research Center for Sustainable Energy Systems (IRC-SES), the SDAIA-KFUPM Joint Research Center for Artificial Intelligence, and the Electrical Engineering Department, King Fahd University of Petroleum \& Minerals Dhahran, 31261, Saudi Arabia.

}
}

\author{Zeyad M. Manaa,
Ayman M. Abdallah,
Mohammad A. Abido,
and Syed S. Azhar Ali
}
\maketitle
\begin{abstract}

Quadrotor systems are common and beneficial for many fields, but their intricate behavior often makes it challenging to design effective and optimal control strategies. Some traditional approaches to nonlinear control often rely on local linearizations or complex nonlinear models, which can be inaccurate or computationally expensive. We present a data-driven approach to identify the dynamics of a given quadrotor system using Koopman operator theory. Koopman theory offers a framework for representing nonlinear dynamics as linear operators acting on observable functions of the state space. This allows to approximate nonlinear systems with globally linear models in a higher dimensional space, which can be analyzed and controlled using standard linear optimal control techniques. We leverage the method of extended dynamic mode decomposition (EDMD) to identify Koopman operator from data with total least squares. We demonstrate that the identified model can be stabilized and controllable by designing a controller using linear quadratic regulator (LQR).    
\end{abstract}

\section{Introduction}
Linear control theory is ideally adapted to creating interpretable control frameworks through investigation of the spectrum components of the related dynamical system. For example, system stability may be determined with the use of spectral analysis \cite{michiels2014stability}. Due to the dynamics' lack of a linear development in nature, application to non-linear systems is unsuitable, which leads to sub-optimal control applications \cite{komeno2022deep}.\\
\indent When the goal is to achieve optimal task performance while abiding by state, actuation, and computing constraints, designing efficient control for dynamic systems remains a difficult task. It is necessary to choose a nonlinear model in order to capture the typical nonlinear behaviors of the majority of dynamical systems. Although global convergence is not guaranteed \cite{boyd2004convex}, solving a nonlinear, non-convex optimization problem is typically required to identify a nonlinear model from data \cite{bruder2019nonlinear}.\\
\indent The majority of nonlinear system identification techniques demand the manual initialization and fine-tuning of training parameters, which has an unclear effect on the model that is produced. For instance, a neural network might be able to model the nonlinear behavior of a dynamical systems \cite{gillespie2018learning}; however, the accuracy of the model depends on the number of hidden layers, the number of nodes used in each layer, the activation function, and the termination condition, all of which must be chosen through trial and error until satisfactory results are obtained. 
Also, the recent development in optimization techniques made it possible for methods like non-linear model predicative control in real-time by \cite{gros2020linear, kouzoupis2018recent} to be implemented. However, these advancements still require sufficiently accurate mathematical models and cannot handle model uncertainty. In order to address some of these problems, researchers focused on data-driven techniques, and learning-based techniques to identify the system's underlying model \cite{hou2013model, hewing2019cautious, han2020deep, wang2021deep, lusch2018deep}. For example \cite{manaa2024data} employed the sparse identification of nonlinear dynamics algorithm to identify the equations of motion of quadrotors, which allows a way to identify the varying parameters in online framework. Jiahao et al. \cite{jiahao2023online} introduced the use of knowledge-based neural ordinary differential (KNDOE) as a dynamic model with usage of a method inspired from transfer learning to improve the system performance in quadrotor platforms. However, these methods require a lot of computation.\\
\indent Contrarily, since linear models can be identified using linear regression, linear model identification does not have the typical drawbacks of nonlinear identification. However, because most of dynamical systems exhibit distinctly nonlinear behavior, linear models are ill-equipped to represent that behavior \cite{bruder2019nonlinear}. Data-driven techniques for linear model approximations also have been researched by \cite{huang2003neural}. However, limitations exist for such data-driven modeling of the dynamical systems. These linear models are locally linearized, thereby they are not capturing the important nonlinear physical characteristics of the system.\\
\indent Recent works have been focused to get a \emph{globally linearized dynamics models} via \emph{Koopman theory} \cite{koopman1931hamiltonian}. Koopman operator governs the advancement of the dynamics (a set of observable functions), which can be interpreted as nonlinear measurement functions of the system states. These results in a linear but generally infinite dimensional representation of the nonlinear dynamics \citep{budivsic2012applied, brunton2022data, lan2013linearization, arbabi2017ergodic}. An approximation for the linear infinite dimensional Koopman operator can be obtained from Dynamic Mode Decomposition (DMD) as in ref. \cite{schmid2010dynamic}. An extension for the DMD algorithm has been carried out by \cite{proctor2016dynamic} to deal with controlled systems. Williams et al. \cite{williams2015data}
further extended the DMD to a version in which snapshots of nonlinear observables can be augmented with the system original states to obtain a \emph{lifted} finite-dimensional approximation of the Koopman operator to account for complicated nonlinearities. Recently, Korda and Mezi{\'c} \cite{korda2018linear} extended the EDMD for controlled dynamical systems using linear control strategies as Model Predictive Control. 
Also, the noisy and biased measurements can be a crucial factor on the identified parameters. Recent research effort has dealt with such a problem \cite{hemati2017biasing, dawson2016characterizing, askham2017robust, abolmasoumi2022robust}. \\
\indent The framework has been used widely in robotics \cite{bruder2021koopman, mamakoukas2021derivative, haggerty2020modeling, han2021desko, chen2022offset, zhu2022koopman} with different variations in designing the observable functions and control design, in power grids \cite{hossain2023data}, in fluid dynamics \cite{markmann2024koopman}, in epidemiology \cite{mezic2024koopman}, and many other fields. Nevertheless, because of the complexity and the topological nature of quadrotors, it is challenging to extend such applications to quadrotors. \\
\indent Folkestad et al. \cite{folkestad2020data} then introduced and used Koopman Eigen function Dynamic Mode Decomposition for such purpose, and more specially to learn the nonlinear ground effect to improve the quadrotors landing performance. Additionally, recent work on quadrotors has been tested by \cite{folkestad2021quadrotor, folkestad2021koopman}. However these methods are complicated, slow, and in many cases, they are applied in low dimensional quadrotors (e.g., planar quadrotor).\\
\indent So, we aim to introduce a simple LQR-based controller for 6 degree-of freedom quadrotors. We will leverage simulation data to build a set of observable functions from our background knowledge about the system and combine it with previously proposed set in literature. We then use the calculated set of observable functions to learn a globally linear model of the unerlying nonlinear quadrotor dynamics using EDMD to approximate the Koopman operator to be used for the LQR design.

\section{Methodology}
\subsection{Koopman Operator Theory and Dynamic Mode Decomposition}

\begin{definition}[Koopman Operator]
Consider the discrete time dynamical system:
\begin{equation} \label{eq:dynamics}
    x^{+}_{k} = f(x_{k}, u_{k}),
\end{equation}
where $x_{k} \in \mathbb{R}^n$ is the state vector, $u_{k} \in \mathbb{R}^l$ is the control input, $f$ is a transition map, and $x^{+}$ is the successor state. The Koopman operator $\mathcal{K}_t$ is an infinite dimensional operator:
\begin{equation*}\label{eq:koop_def}
\mathcal{K}_t \xi = \xi \circ f(x_k, u_k),
\end{equation*}
acting on $\xi \in \mathcal{H}: \mathbb{R}^n \times \mathbb{R}^l \mapsto \mathbb{R}$, where $\circ$ denotes function composition. \hfill $\Box$
\end{definition}

\begin{figure}[]
    \centering
    \includegraphics[width=0.5\textwidth]{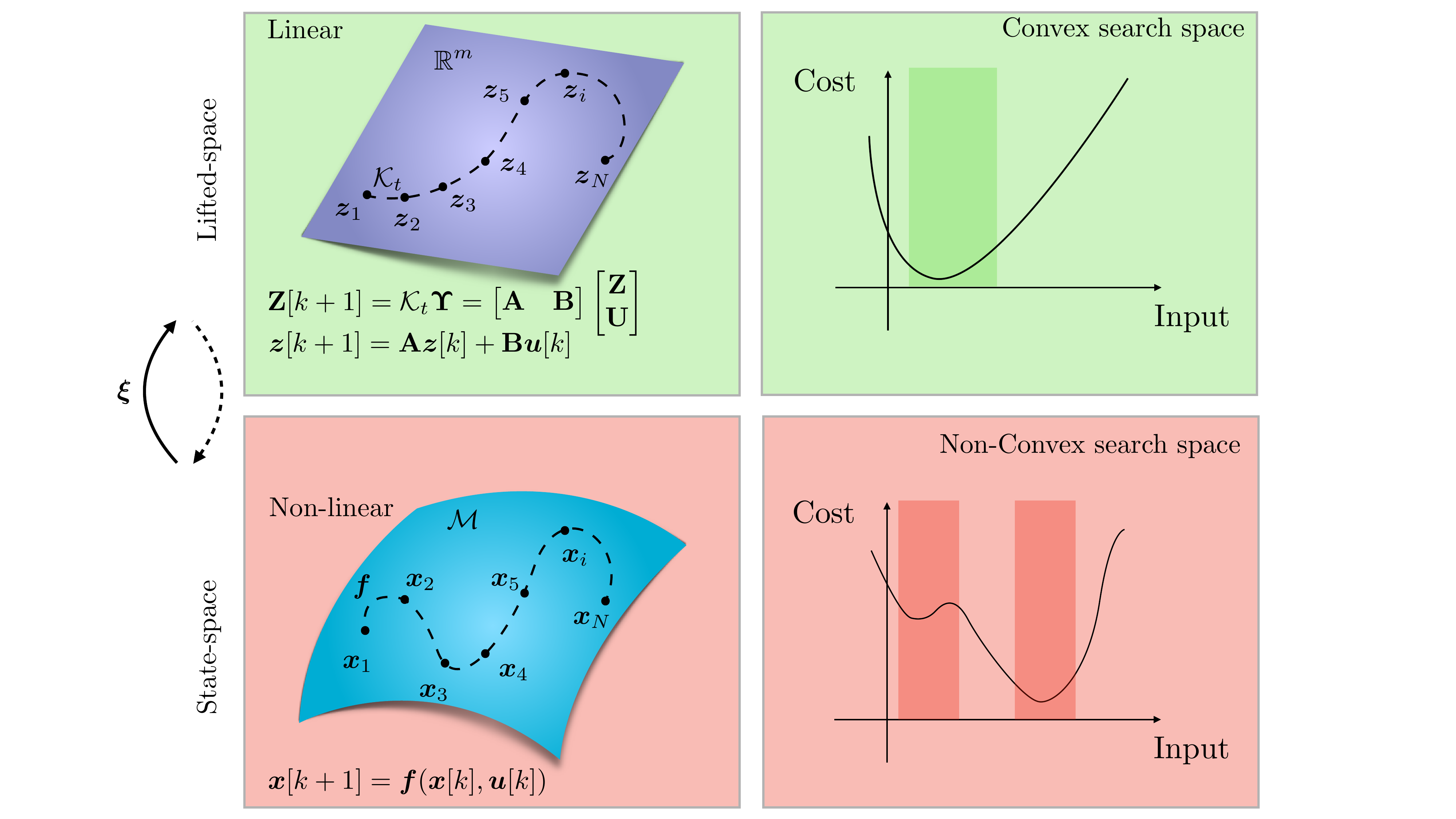}
    \caption{Illustration of the Koopman Operator: The Koopman Operator ($\mathcal{K}_t$) maps state-space into a lifted linear space governed by ${\xi}$. Green panels represent linear space (convex search space for control design), while red panels represent nonlinear state-space (non-convex search space).}
    \label{fig:koopman_illustration}
\end{figure}

The Koopman operator provides a linear representation of a nonlinear system in infinite-dimensional space by acting linearly on the Hilbert space $\mathcal{H}$ of measurement functions $\xi$. A finite representation of the Koopman operator is sought in practice, however this transformation exchanges nonlinear, finite-dimensional dynamics for linear, infinite-dimensional dynamics. An advantage of linear systems in control theory is the convex search space for controllers like LQR and MPC (see Fig. \ref{fig:koopman_illustration}).

In order to get around the problems caused by infinite dimensionality, we use data to approximate the Koopman operator. Think of a system whose inputs and states have been recorded as
\begin{align*}
    \mathcal{D} := \{x_k, u_k: k \in [0, (T-1)T_s]\cap\mathbb{Z}_{0}\},
\end{align*}

\begin{assumption} \label{ass:D}
    The set $\mathcal{D}$ exists. \hfill $\Box$
\end{assumption}

Let assumption \ref{ass:D} hold. Define:
\begin{subequations}
\begin{align}
    &\Gamma := \mat{u(0)~ u(T_s)~ \dots ~ u((T-1)T_s)} \in \mathbb{R}^{l \times T},\\
    &X := \mat{x(0)~ x(T_s)~ \dots~ x((T-1)T_s)} \in \mathbb{R}^{n \times T},\\
    &X^{+} := \mat{x(1)~ x(T_s)~ \dots~ x((T)T_s)} \in \mathbb{R}^{n \times T}.
\end{align}
\end{subequations}

\begin{assumption} \label{ass:PE}
    Considering $T \geq n+l$, the matrix $\mat{X\\ \Gamma}$ has full row rank. \hfill $\Box$
\end{assumption}

\begin{definition}[Regression using DMD] \label{def:DMD}
    Consider a dynamical system ${x}^{+}_{k} = f(x_k, u_k) \approx A x_k + B u_k$ and dataset $\mathcal{D}$. The system can be written as:
    \begin{align} 
        X^{+} &= A X + B \Gamma \\
        &= \begin{bmatrix} A & B \end{bmatrix} \begin{bmatrix} X \\ \Gamma \end{bmatrix} = \hat{\mathcal{K}_t} \Omega.
        \label{eq:dmdc}
    \end{align}
    Assuming \ref{ass:PE} holds, solve:
    \begin{align}
        \label{eq:soln_ls}
        \hat{\mathcal{K}_t} = \argmin_{\hat{\mathcal{K}_t}} \|X^{+} - \hat{\mathcal{K}_t} \Omega\|_{\text{F}} = X^{+} \Omega^{\dagger},
    \end{align}
    where $\dagger$ denotes the Moore-Penrose pseudo-inverse, and $\|\cdot\|_{\text{F}}$ is the Frobenius norm. \hfill $\Box$
\end{definition}

\begin{remark}
    The Moore-Penrose pseudo-inverse is calculated using SVD. \hfill $\Box$
\end{remark}

The original DMD algorithm was developed for non-controlled systems \cite{schmid2010dynamic}. In \cite{proctor2016dynamic}, it was extended for controlled settings. For more on DMD, see \cite{schmid2010dynamic, tu2013dynamic}.

DMD can be extended using EDMD by lifting states into a higher dimensional space with observables. The main difference between nominal DMD and EDMD is that in nominal DMD, the observables are identity maps. EDMD uses selected observables to approximate the Koopman operator. These observables can be found by trial and error or system knowledge. For more on this, see \cite{li2017extended, otto2019linearly, yeung2019learning, lusch2018deep}. Instead of just system states, consider additional observables:
${\Xi}({x}) = \begin{bmatrix}
    \xi_1({x}) & \xi_2({x}) & \dots & \xi_p({x})
\end{bmatrix}^{\top}$, such that the system (\ref{eq:dmdc}) becomes
\begin{align} 
    \Xi(X^{+}) = \begin{bmatrix} A & B \end{bmatrix} \begin{bmatrix} \Xi(X) \\ \Gamma \end{bmatrix} = \hat{\mathcal{K}_t} \bar{\Omega},
    \label{eq:edmdc}
\end{align}
solved similarly to (\ref{eq:soln_ls}) by
\begin{align}
    \hat{\mathcal{K}_t} = \argmin_{\hat{\mathcal{K}_t}} \|\Xi(X^{+}) - \hat{\mathcal{K}_t} \bar{\Omega}\|_{\text{F}}.
    \label{eqn:normal_least_squares}
\end{align}

\begin{remark}
    If lifting functions are identity, EDMD reduces to nominal DMD. \hfill $\Box$
\end{remark}
The linear lifted approximation of nonlinear dynamics in (\ref{eq:dynamics}) is
\begin{align*}
    z_{k}^{+} &= Az_{k} + Bu_{k} \\
    x_k &= Cz_{k}.
\end{align*}
The implemented algorithm is summarized in Algorithm \ref{alg:koopman-identification}.
\begin{algorithm}[]
\caption{Koopman Identification using Total-Least-Squares with $L_1$ Regularization}
\label{alg:koopman-identification}
\begin{algorithmic}[1]
\REQUIRE Dataset $\mathcal{D}$
\STATE  Lift data via (\ref{eqn:lift})
\STATE Get the approximated Koopman operator $\hat{\mathcal{K}}_t$ via (\ref{eqn:normal_least_squares})
\STATE  Extract the identified linear matrices $A$, $B$ from $\hat{\mathcal{K}}_t$
\RETURN $\hat{\mathcal{K}}_t$, $A$, and $B$
\end{algorithmic}
\end{algorithm}

\subsection{Quadcopter dynamics}
Quadrotor is assumed to be a rigid body with a 6 degrees of freedom with a mass $m$ and diagonal inertia tensor of ${J} = \text{diag(} J_{xx}, J_{yy}, J_{zz}\text{)}$. Our model is built upon  quaternion formulation for attitude parameterization and is similar to the standard models found in \cite{torrente2021data, carino2015quadrotor}. For better interpretation of the results quaternions are converted to Euler angles. The state space then is 13-dimensional or 12-dimensional respectively space, based on either quaternion, or Euler angles formulation (respectively). Suppose quaternions are used for attitude formulation, so $x \in \mathbb{R}^{13}$ and is governed by the dynamics
\begin{align}
\label{eq:3d_quad_dynamics}
\dot{{x}} =
\deriv{}{t}\begin{bmatrix}
{{p}}_{WB} \\ 
\dot{{p}}_{WB} \\
{{q}}_{WB} \\
{\omega}_B
\end{bmatrix} = 
{f}({x}, {u}) =
\begin{bmatrix}
{p}_W \\ 
\frac{1}{m}\;{q}_{WB} \cdot {T}_B + {g}_W \\
\frac{1}{2} {q}_{WB} \otimes {\omega_B} \\
{J}^{-1}\left({\tau}_B - \omega_B \times {J}\omega_B\right)
\end{bmatrix},
\end{align}
where ${T}_B$ is the summation of quadrotor's thrust, ${g}_W= [0, 0, \SI{-9.81}{\meter\per\second^2}]^\top$ is Earth's gravity, ${\tau}_B$ is the torque, ${p}_{WB}$, position vector from world to body frame, and $q_{WB}$  quaternion from world to body frame. Let $\otimes$ denotes the quaternion-vector product by , where, for example, $p\otimes q = qp{q}^{*}$, and $q^{*}$ is the quaternion's conjugate. 
\begin{align}
{T}_B =  \mat{0 \\ 0 \\ \sum T_i} \quad
\text{and} \quad 
{\tau}_B = 
\mat{l (-T_0 - T_1 + T_2 + T_3) \\
l (-T_0 + T_1 + T_2 - T_3) \\
c_\tau (-T_0 + T_1 - T_2 + T_3)},
\end{align}
where $l$ is the distance from
the center of mass to the motor axis of action and $c_\tau$ is a constant for the rotor's drag torque.

The famous Runge-Kutta fourth order scheme will be considered to integrate the dynamics ${\dot{x}}$ in a discrete manner with a time step of $\Delta t$
\begin{equation}
\label{eq:discretized_nominal_dynamics}
{{x}_{k+1} = {x}_{k}} + \int_{k\Delta t}^{(k+1)\Delta t} {f}({x}, {u}, \Delta t) d \tau
\end{equation}
\subsection{Data Collection and Koopman Training}
A PD controller is designed to track the generated trajectories of the quadrotor. Five helical trajectories are simulated with quadrotor's parameters in table \ref{tab:physical}, providing a diverse set of scenarios for analysis. 
\begin{table}[]
    \centering
    \caption{Physical parameters of the quadrotor}
    \label{tab:physical}
    \begin{tabular}{cc}
    \toprule
    \textbf{Parameter} & \textbf{Value} \\
      \midrule
      Mass $m$ (kg) & 0.18 \\
      Gravity $g \cdot \hat{{k}}$ (m/s\textsuperscript{2}) & 9.81 \\
      Inertia ${J}$ (kg$\cdot$m\textsuperscript{2}) & $\begin{bmatrix}
        0.00025 & 0 & 0 \\
        0 & 0.000232 & 0 \\
        0 & 0 & 0.0003738
      \end{bmatrix}$ \\
      Moment arm $l$ (m) & 0.086 \\
      \bottomrule
    \end{tabular}
\end{table}
A time step $\Delta t$ of $0.01$ seconds is set in the simulation to ensure detailed temporal resolution for 30 secendos for each trajectory. Random parameters are utilized to also introduce variability in the trajectories. The helix radius varies randomly between 1 and 5 meters, while the height experiences random fluctuations within the range of 1 to 6 units.

We trained our EDMD model with total least squares optimizer dealing with the data using a set of observable from our background of the system and we augment it with another observable functions retrieved from literary sources (see ref. \cite{chen2022koopman}) defined as 
\begin{equation} \label{eqn:lift}
    \begin{aligned}
        {\Xi}({x}) = \big[&1,~ x,~ p_{WB},~ \dot{p}_{WB},~ \sin(p_{WB}),~ \dots \\ & \cos(p_{WB}),~ \text{vec}(R \times \omega_{WB})
    \big],
    \end{aligned}
\end{equation}
where $\text{vec}(*): \mathbb{R}^{n\times n} \mapsto \mathbb{R}^{n^2 \times 1}$ is an operator that maps a matrix into a vector by stacking the columns of the matrix, and ${R}$ is the rotation matrix representing the quadrotor attitude. 

\subsection{Linear Quadratic Regulator Meets Koopman Linearization}
\begin{assumption} \label{ass:cont_obs}
    The identified pairs $(A, B)$ and $(A, C)$ are controllable and observable respectively. \hfill $\Box$
\end{assumption}
\begin{remark}
    Assumption \ref{ass:cont_obs} can be numerically verified for the identified system's matrices. \hfill $\Box$
\end{remark}
Now, consider a quadratic cost function:
\begin{align} \label{eqn:cost}
    \mathcal{J} = \minimize_{u_0,\dots,u_{N-1}} \sum_0^{N-1} x^{\top}(\tau)Qx(\tau) + u^{\top}(\tau)Ru(\tau) d\tau,
\end{align}
where $Q = Q^{\top} \succeq 0$ is a weight matrix for the cost of deviation of the state $x$ from the reference point, and $R = R^{\top} \succ 0$ weighs the cost of control action. If a linearized version of (\ref{eq:dynamics}) is considered, it is possible to have a matrix $L$, and a control law in the form of $u = -Lx$ such that the cost $\mathcal{J}$ is minimum. However, this controller is linear, and in fact, it will be optimal only around the neighbourhood of the fixed point at which the linearization took place. \\
\indent Here, instead of having a local linear model for our system, we will instead use the global linear model derived from Koopman approximation. The cost function still holds but with minor modifications as follows
\begin{align*}
    \bar{Q} = \mat{Q_{n \times n} & 0_{p-n \times p-n}\\
    0_{p-n \times p-n} & 0_{p-n \times p-n}}_{p \times p}, \quad \bar{R} = R
\end{align*}
So, the cost function given in (\ref{eqn:cost}) will remain the same, while replacing $x$ by $z$, and $Q$ by $\bar{Q}$.
\begin{remark}
    In this formulation, the matrix $R$ stay as it is as control inputs are remain un-lifted. \hfill $\Box$
\end{remark}
In that sense, an optimal LQR controller can be designed for our koopman linearization of our system. \\
\indent Let $Q = \mathbb{I}_{12 \times 12} \times 10^{3}$, and $R = \mathbb{I}_{4 \times 4}$ be chosen for controller design. 

\begin{figure}[t]
     \centering
     \begin{subfigure}[b]{0.4\textwidth}
         \centering
         \includegraphics[width=0.8\textwidth]{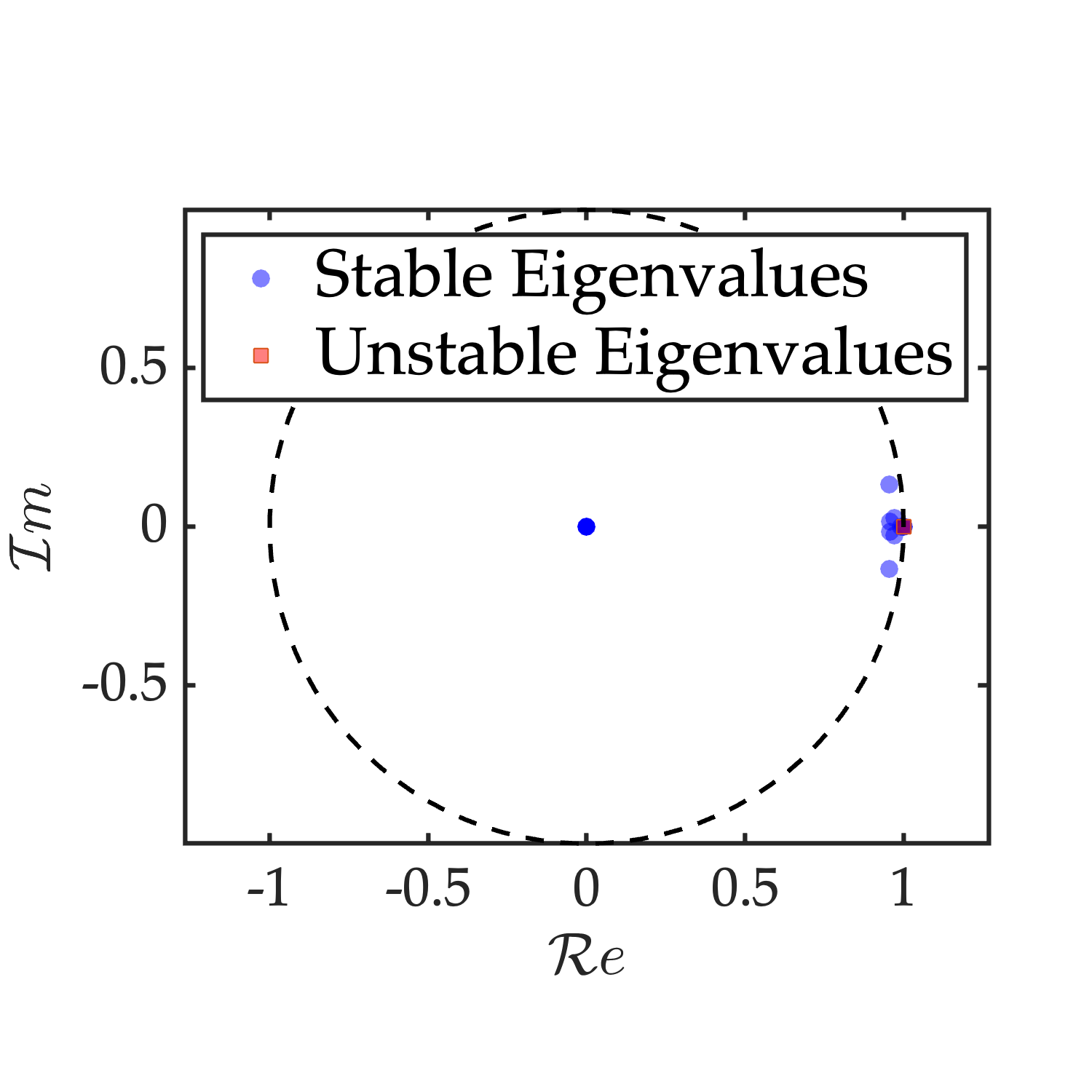}
         \caption{Uncontrolled dynamics eigenvalues}
         \label{fig:sectrum_uncont}
     \end{subfigure}
     \hfill
     \begin{subfigure}[b]{0.4\textwidth}
         \centering
         \includegraphics[width=0.8\textwidth]{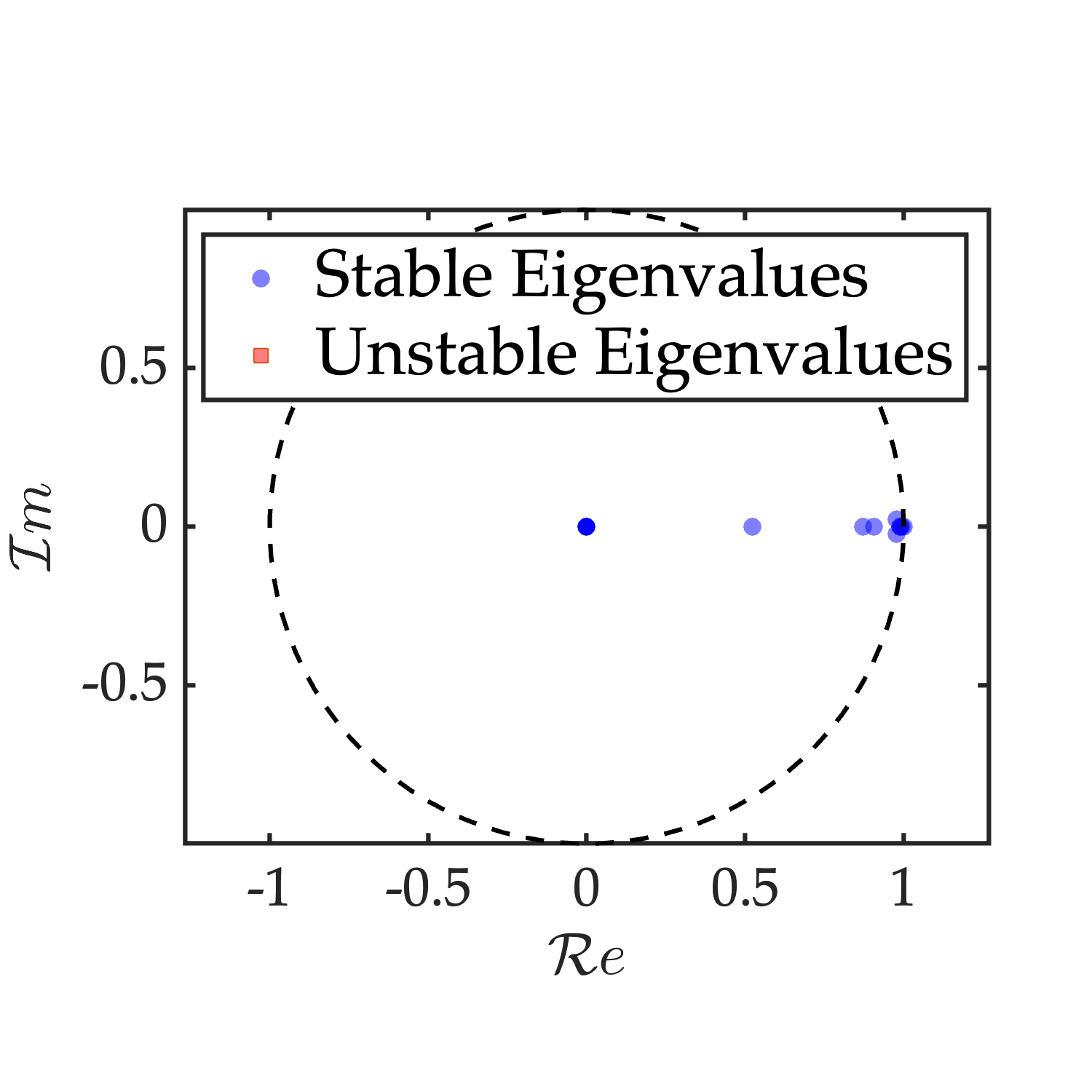}
         \caption{Controlled dynamics eigenvalues}
         \label{fig:spectrum_cont}
     \end{subfigure}
        \caption{The spectrum analysis of the discovered system without control (right) with LQR stabilization (left).}
        \label{fig:spectrum}
\end{figure}

The system stability is checked in the lifted space by checking the spectrum of the matrices ${A}$ and $(A - BK)$ and got the results in Fig. \ref{fig:spectrum}. Even in the lifted space we got a confirmation of the system instability with some of the eigenvalues outside the unit disk Fig. \ref{fig:sectrum_uncont}. The LQR design done on the lifted space rendered the matrix $(A - BK)$ Hurwitz as seen in Fig. \ref{fig:spectrum_cont}. 

Furthermore, an excellent performance is noted for Koopman control as compared to the nominal PID in the rotational domain of the quadrotor (i.e. the Euler angles and their rates). This is can be seen clearly from Table \ref{tab:NRMSE_ALL} for the case of Euler angles, as the exhibit the lowest \%RMSE. Although the average \%RMSE in Table \ref{tab:NRMSE_ALL} between the nominal PID and Koopman in terms of angular velocities is quite high, the Koopman controller a converging behaviour to the reference value.

\begin{figure*}[]
    \centering
    \includegraphics[width=0.95\textwidth]{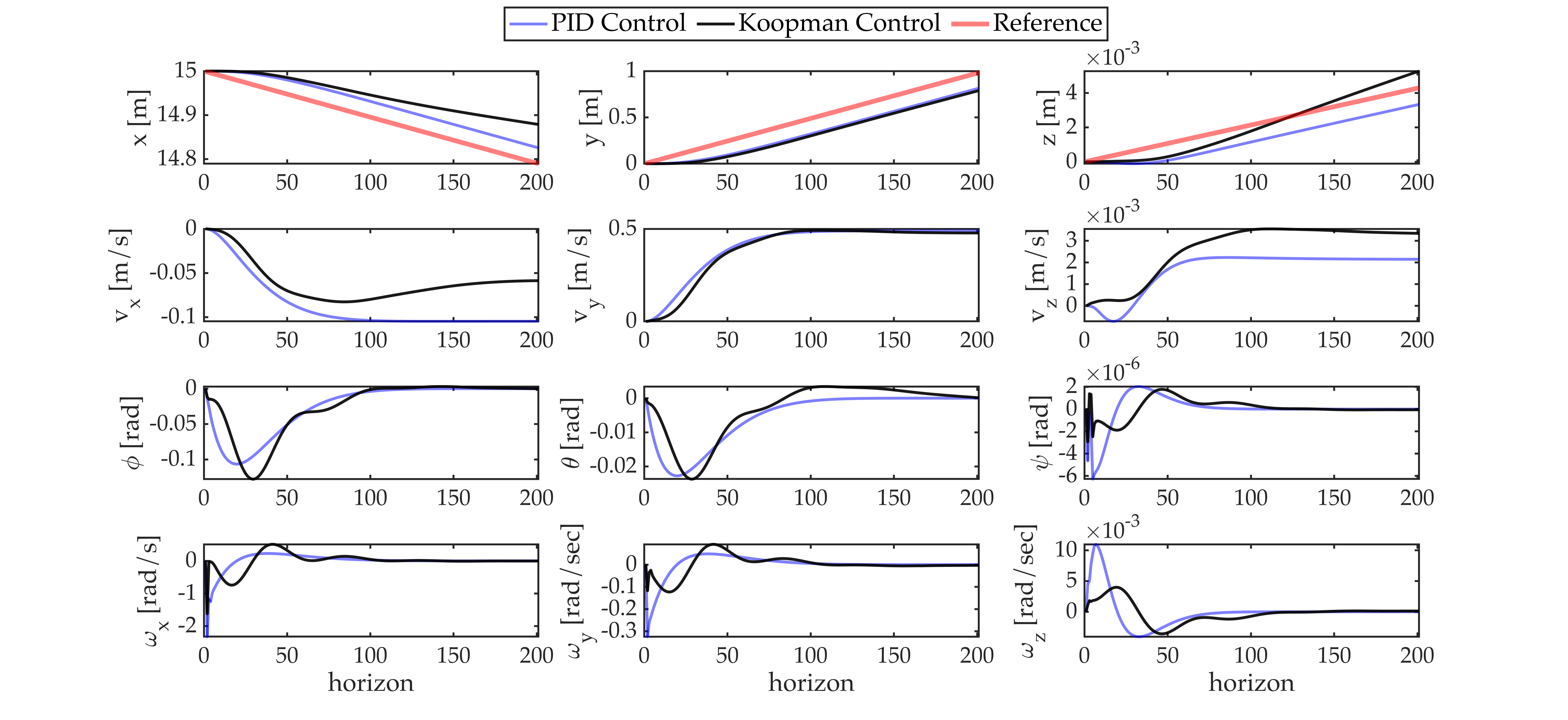}
    \caption{Illustration of the performance comparison between Nominal PID Control (blue), Koopman Control (black), and the Reference trajectory (red) for a quadrotor. The plots depict the state trajectories over a 200-step prediction horizon, covering the states of the quadrotor including position ($x$, $y$, $z$), velocity ($\dot{x}$, $\dot{y}$, $\dot{z}$), Euler angles ($\phi$, $\theta$, $\psi$), and angular velocities ($\omega_x$, $\omega_y$, $\omega_z$).}
    \label{fig:prediction}
\end{figure*}

\section{Results and Discussions}
In addition to the designed LQR controller, we designed a typical PID controller compiled from several sources as in \cite{johnson2017control, luukkonen2011modelling, mahony2012multirotor} to serve as a typical benchmark with the new method.

To evaluate the performance of the learned predictors, we considered the normalized Root Mean
Square Error (NRMSE) as a system metric.
\begin{equation}
    \text{NRME} = 100 \times \sqrt{\dfrac{ ||{x}_{\text{pred}} - {x}_{\text{true}}||_{2}^{2}}{||{x}_{\text{true}}||^{2}_{2}}}
\end{equation}
The average NRMSE of a simulated case that take place for 1.5 second, can be viewed in table \ref{tab:NRMSE_ALL}.  

\begin{table}[]
\centering
\caption{The average NRMSE of the states between the Koopman and PID based-controllers}
\label{tab:NRMSE_ALL}
\begin{tabular}{@{}ll@{}}
\toprule
{states}        & {\%RMSE}     \\ \midrule
${{p}}_{WB}$ -- Position     & $3.2529 \pm 2.3216$ \\
$\dot{{p}}_{WB}$ -- Velocity & $4.8129 \pm 3.8608$ \\
${{\mathcal{E}}}_{WB}$ -- Euler angles       & $2.4398 \pm 2.4263$ \\
${{\omega}}_{B}$ -- Angular velocity    & $7.8525 \pm 6.6367$ \\
\hline
{Mean}     & $4.5895 \pm 3.8114$ \\ \bottomrule
\end{tabular}%
\end{table}

The results are within a good range for being used in control purposes. The fact that these results are coming from a globally linearized model for a highly nonlinear model such as the quadrotor is remarkable. The predictive capability of the discovered linear mode is also remarkable. The prediction capability is depicted in Fig. \ref{fig:prediction} for pre-calculated pd control inputs. The linear model operated in a good way which validate the discovered model.

\section{Conclusion}
To sum up, this research offers a data-driven method for creating efficient and ideal control schemes for quadrotor systems. Globally linear models in higher dimensional space can identify and approximate the dynamics of a quadrotor by utilizing Koopman operator theory and EDMD. With the use of a LQR, the found model enables the design of a stabilizing controller and the prediction of quadrotor dynamics. This strategy provides a more straightforward and understandable framework for control design while overcoming the drawbacks of conventional nonlinear control techniques. The suggested approach creates opportunities for open research in modeling quadrotors using global linear models.

Although the algorithm's promising performance, some limitations we have noticed should be highlighted. First, the system does not scale on different type of trajectories other than the learnt trajectories for our settings limiting its ability to account for general purpose model. When we augmented the data with different type of trajectories, the learning process became harder. Also, we have found that our identified model is good on small control horizon (i.e. $\sim 150$ step, equivalent to $\sim 1.5$ sec). 

Nevertheless, we here have laid out one of the initial steps to deal with such problems for complicated system with hard topology like quadrotors. For example, the problem of scalability could be approached by building large data sets with various type of trajectories and conduct ensemble learning on the system. Also, the problem of the limited prediction horizon can be tackled using a recursive estimate of the Koopman operator every $N$ period. Those potentials open the door to myriad of research to prove such concepts.


{\small
\bibliographystyle{ieeetr}
\bibliography{ref}
}
\end{document}